# Monitoring, Analyzing, and Controlling Internet-Scale Systems with ACME

David Oppenheimer, Vitaliy Vatkovskiy, Hakim Weatherspoon, Jason Lee[†], David A. Patterson, and John Kubiatowicz

University of California, Berkeley,   [†]University of California, Los Angeles
{davidopp, vatkov, hweather, jlee81, pattrsn, kubitron}@cs.berkeley.edu

**Abstract**

*Analyzing and controlling large distributed services under a wide range of conditions is difficult. Yet these capabilities are essential to a number of important development and operational tasks such as benchmarking, testing, and system management. To facilitate these tasks, we have built the Application Control and Monitoring Environment (ACME), a scalable, flexible infrastructure for monitoring, analyzing, and controlling Internet-scale systems. ACME consists of two parts. ISING, the Internet Sensor In-Network agGregator, queries "sensors" and aggregates the results as they are routed through an overlay network. ENTRIE, the ENgine for TRiggering Internet Events, uses the data streams supplied by ISING, in combination with a user's XML configuration file, to trigger "actuators" such as killing processes during a robustness benchmark or paging a system administrator when predefined anomalous conditions are observed. In this paper we describe the design, implementation, and evaluation of ACME and its constituent parts. We find that for a 512-node system running atop an emulated Internet topology, ISING's use of in-network aggregation can reduce end-to-end query-response latency by more than 50% compared to using either direct network connections or the same overlay network without aggregation. We also find that an untuned implementation of ACME can invoke an actuator on one or all nodes in response to a discrete or aggregate event in less than four seconds, and we illustrate ACME's applicability to concrete benchmarking and monitoring scenarios.*

## 1. Introduction

Consider the following scenarios:

(1) **Benchmarking:** You've just written what you're sure is the world's fastest, most reliable distributed hash table, and you want to see how it stacks up against other DHTs by measuring its performance and robustness under stressful scenarios such as high load, large groups of nodes suddenly dying, or nodes quickly joining and leaving.

(2) **Testing:** You've just implemented a complicated byzantine agreement protocol for Internet-scale systems. You want to make sure it can truly handle up to the specified number of failures, and can handle them occurring when the protocol is in various states.

(3) **System management:** You're in charge of a large, globally distributed network service testbed. As a conference deadline approaches, researchers occasionally run buggy prototype software that causes severe shortages of CPU, memory, network bandwidth, and disk space. You decide to implement the following policy: if a user is observed to be using "too much" CPU time, memory, or disk space, kill all processes the user is running; if the user is using too much network bandwidth, place a cap on the bandwidth the user is allowed; and in either case, send email to the user. If the condition reappears after a short time, repeat the process and also attempt to delete cron jobs that might be automatically restarting the offending processes.

The most common way to implement these tasks today is to custom write hundreds or thousands of lines of code that execute the desired monitoring and control policy. While some existing systems enable policy-driven monitoring of large distributed systems, and a few tools can introduce controlled events during such monitoring, we believe a single system can provide sufficient expressiveness of configuration for all three classes of tasks while targeting large-scale applications that are geographically distributed across the Internet.

The primary challenges in building such a system are

(i) **Scalability**: The monitoring component must be capable of collecting data from hundreds or thousands of nodes

(ii) **Flexibility**: The rule engine must be easily configurable to specify a wide range of monitoring conditions and control actions to be taken when the monitoring conditions are met. Also, it must be easy to add new application-level data sources and control actions.

(iii) **Robustness**: The system must handle failures in the managed, or managing, application: the monitoring component must report approximate results when some nodes fail or become partitioned, and redundant monitoring components should be usable by the control component, in case the monitoring component itself fails.

As research into "Internet scale" systems gains steam, there is an increasing need for tools to benchmark, test, monitor, and control these systems, both before and after deployment. In this paper we describe the design and implementation of ACME, the Application Control and Monitoring Environment, a scalable, flexible infrastruc-ture that can perform all of these tasks and that meets the aforementioned challenges.

ACME is built from two principal parts. ENTRIE, the ENgine for TRiggering Internet Events, is a user-configured trigger engine that invokes "actuators" in response to conditions over metrics collected from "sensors." This sensor data may come directly from nodes or from ISING, the Internet Sensor in-Network aGgregator, which is the second part of our infrastructure. ISING is a very simple distributed query processor for continuous queries over sensor data streams; it broadcasts queries to sensors using a tree-based overlay network and then collects and aggregates resulting data streams as they travel back up through the network. ISING trades off expressiveness for ease of implementation by using its own query language rather than SQL. ISING is built on top of QTree, a spanning tree overlay network with a configurable topology that is used by ISING for query distribution and result aggregation.

ACME meets the challenges we have mentioned as follows. To achieve **scalability**, ISING broadcasts queries and collects results using a peer-to-peer overlay network, and it aggregates results as they travel through the network. In Section 4.2 we show that this aggregation is quite beneficial. For **flexibility**, ENTRIE allows users to specify trigger conditions and their corresponding actions using an XML configuration file. Also, we have implemented standards-compliant "sensors," as well as "actuators," to demonstrate the ease with which new *application-level* sources of monitoring data and sinks for control actions can be added to the system. Section 3.4 shows sample configuration files for benchmarking and system management, and Section 4.3 shows the application-level sensors and actuators in use during a benchmark of two structured peer-to-peer overlay networks.[1] Finally, for **robustness**, ISING uses timeouts to deliver a node's aggregated result up the tree if the node does not hear from all of its children in a timely fashion.

The remainder of this paper is organized as follows. In Section 2 we provide some background on the sensor and actuator metaphor, and their implementation in the system we describe. Section 3 describes the design and implementation of ACME, including the ISING data collection infrastructure and the ENTRIE trigger engine. In Section 4 we evaluate ISING and ACME as a whole and demonstrate ACME being used in a benchmarking scenario. In Section 5 we discuss related work, Section 6 describes future work including plans for deploying ISING on PlanetLab, and in Section 7 we conclude.

Although we leave a discussion of related work to the end of the paper, we wish to mention at this juncture that ACME bears a strong resemblance to three existing systems. Sophia [29] is a distributed expression evaluator for Prolog statements over sensors and actuators; it is in some ways a more general purpose version of ACME. PIER [11] is a distributed SQL query engine for stored data and Internet sensors; it is thus a more general purpose version of ISING. Finally, TinyDB [15] is a distributed SQL query engine for wireless sensors; it bears an even stronger resemblance to ISING in that it performs in-network aggregation when responding to queries. We feel that ACME complements these projects by exploring a distinct design point that offers its own unique lessons.

## 2. Sensors and actuators

Because ACME's primary capabilities are the ability to aggregate streaming sensor data in real time and to control system operation via "actuators," we briefly provide some background on these metaphors, existing implementations of them, and the new sensors and actuators that we wrote.

Although the sensor/actuator metaphor for observing and controlling distributed systems is more than a decade old [16], the sensor side of this equation has recently received increased attention due to its incorporation as a fundamental building block of the PlanetLab testbed [19]. A PlanetLab sensor is an abstract source of information derived from a local node [23]. Sensor data is accessed via a sensor server, which implemented as an HTTP server, that provides access to one or more of the sensors on the node. The sensor server for a particular sensor runs on the same port number across all physical nodes in the system. A sensor can be queried for a value by issuing an HTTP request whose format is described in [23]. The query URL contains the name of the sensor and optional arguments. A sensor returns one or more tuples of untyped data in comma-separated value format. An example of a sensor currently available and that our system uses as a data source is *slicestat*, which provides, for each *slice* (which can be thought of for the purposes of this discussion as a user), various pieces of resource usage information such as the amount of physical memory in use by the slice, the number of tasks executing on behalf of the slice, and the rate of sending and receiving network data over the past 1, 5, and 15 minute intervals.

The PlanetLab sensors that have been developed and deployed to date allow monitoring of operating system and network statistics, such as those described in the previous paragraph. In order to allow controlled and uniform data collection from applications and their log files, we implemented several of our own sensors to provide data about *application* components. The applications we targeted for evaluating ACME were two structured peer-to-peer overlay networks (Chord [25] and Tapestry [33]). For Tapestry we embedded a small HTTP server inside each Tapestry instance; this HTTP server serves as a sensor server for the sensors exported by Tapestry. The sensors we implemented for Tapestry return the number of various types of messages that have passed through a node (*e.g.,* locate object, publish object); the Tapestry instance's routing table; and the latency, bandwidth, and loss statistics for a requested peer (or all peers) as collected by Tapestry's Patchwork background route maintenance component. For both Tapestry and Chord we implemented a log file reader that collects instrumentation and debugging data that is written to disk as the application runs.

---
[1] Although the applications we target for monitoring and controlling in this paper are drawn from the domain of structured peer-to-peer overlay networks, our infrastructure can be easily adapted to work with other types of distributed applications.

In addition to implementing application-level sensors, we have extended the PlanetLab sensor metaphor to include "actuators," an idea that was also recently proposed in [29]. Actuators allow one to *control* entities, much as sensors allow one to *monitor* entities. A program interacts with our actuators in exactly the same way that a program interacts with a sensor: the program sends an HTTP query to a sensor server running on the local host requesting a URL that specifies the name of the actuator and any additional arguments. The actuator returns an acknowledgement that the action was taken or an error message indicating why it was not taken.

Our primary interest in developing actuators is to allow fault injection for robustness benchmarks and tests. Our actuators allow the user to inject perturbations into the environment by starting application processes (and having them join an existing application service such as a distributed hash table), killing nodes, rebooting nodes, and modifying the emulated network topology, through a simple shell wrapper. (The last two features are available only when running on a platform that supports them, *e.g.,* Emulab). We have also embedded actuators into applications themselves, much as we did with sensors. This allows a program to inject a fault into another program using the same interface as it uses to inject faults into the environment. Among the fault injection actuators we have implemented are ones that cause a decentralized routing layer node to drop a fraction of its packets, and to cause a decentralized routing layer workload generator (an instance of which is running in each process of the decentralized routing layer) to change its workload model as the routing layer continues to run. We implemented the first actuator within Tapestry, and the second actuator within both Tapestry and Chord.

In the remainder of this paper, when we refer to a PlanetLab sensor (or actuator), we are referring to a data source (or command sink) that is addressable through the PlanetLab sensor interface. Also, we note that sensors and actuators raise a host of security and protection issues that we do not address in our current implementation.

## 3. ACME design and implementation

In this section we describe ACME and its two principal components: the ENTRIE trigger engine and the ISING sensor aggregator (which is in turn built on top of QTree).

### 3.1. High-level ACME architecture

Figure 1 depicts ACME's high level architecture. The root of the tree is depicted by the boxes drawn above the horizontal line. One representative non-root node is depicted by the boxes below the horizontal line. Each dot is a physical node running the components (boxes) below the horizontal line. Thus the physical nodes form a tree, and all nodes are functionally symmetric, except the root of the tree which additionally runs ENTRIE and stores experiment specifications.

Zeroing in on a single node, a single Java virtual machine (JVM) runs: the SEDA event-driven framework and asynchronous I/O libraries (not shown) [30]; QTree, a configurable overlay network that forms a spanning tree over the nodes in the system; and ISING, a simple distributed query processor specialized for distributing queries to, and aggregating results from, sensors and actuators.

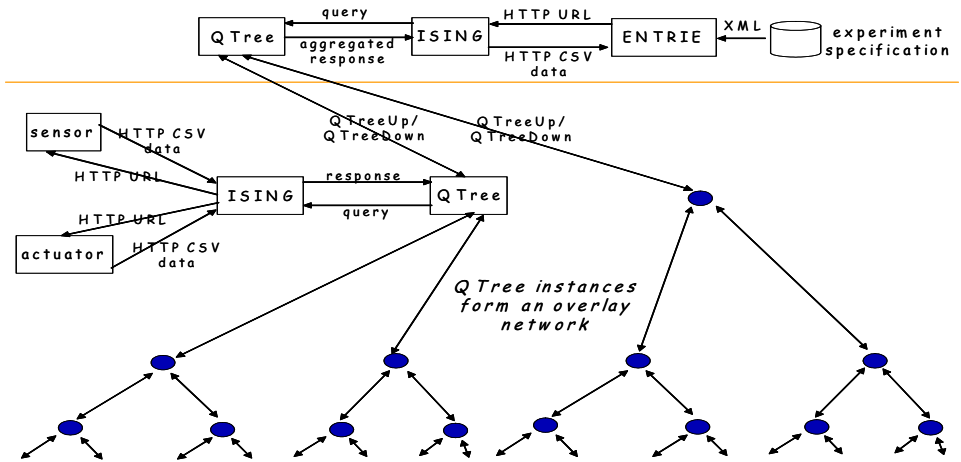

**Figure 1: Overview of the ACME architecture.** A user interacts with ENTRIE running on the root node (drawn above the horizontal line). ENTRIE queries the root ISING instance, which broadcasts the query and aggregates responses using the QTree overlay. The ISING instances running on each node communicate with local sensors and actuators running on those nodes. The sensors and actuators on the root node are omitted from the figure for clarity.

Sensors and actuators run in a separate process from the JVM running SEDA, QTree, and ISING. The root ISING instance itself exports a PlanetLab sensor interface, and it can therefore be used directly as a service. Indeed this is how it is used by ENTRIE, a trigger system that executes user-specified actions when user-specified conditions are met. The conditions are generally ISING queries or timers, and the triggers are generally actuator invocations, but other types of conditions and actions can be specified. ENTRIE runs in a separate process from SEDA/QTree/ISING. All communication among components running in separate processes uses TCP with persistent connections.

### 3.2. QTree

QTree is a configurable spanning tree overlay network that is used by ISING for query distribution and result aggregation. The spanning tree is formed as suggested for aggregation in [3] -- the paths from each node to a designated root form the tree, and aggregation takes place at non-leaf nodes. QTree currently implements three tree topologies: in one (DTREE), the path from a node to the root is a direct TCP connection, and in the two others the path is the overlay routing path the node would use when routing to the root in Tapestry (TTREE) and Chord (CTREE), respectively.

TTREE is formed by following the natural Tapestry routing path from all non-root nodes to the root. As described in [6], this policy ensures that children of nodes near the leaves are close to their parents in terms of network latency, while children of nodes near the root are farther from their parents. This policy is beneficial in an aggregation network because since most edges of the graph are near the leaves, and edges near the leaves are latency-optimized, most data is sent over low-latency links. The smaller number of links near the root, carrying (as we will see) aggregated data, are higher latency. Thus TTREE is beneficial, assuming wide-area network bandwidth is expensive in performance and financial cost.

TTREE is self-organizing, automatically incorporating new nodes as they join the network and remaining fully connected even in the face of failures. Note that QTree does not use Tapestry to route messages; QTree uses Tapestry's initial topology to form the tree and subsequent topology updates (as Tapestry detects nodes departing and joining the network) to re-form the tree, but QTree sends network messages among nodes directly over persistent TCP connections (that are shared with Tapestry).

CTREE is formed by following the natural Chord routing path from all non-root nodes to the root. Due to time constraints we have not yet evaluated ACME using CTREE.

QTree exports a simple interface to applications:

- *NewTree()*: form a new tree rooted at the calling node and return a handle for the tree

- *QTreeDown(tree, message):* send *message* to all descendants of the calling node in *tree*
- *QTreeUp(tree, message):* send *message* to the parent of the calling node in *tree*
- *CountChildren(tree):* return the number of children of the calling node in *tree*
- *WhatsMyLevel(tree):* return the level of the calling node in *tree*

The simplicity of this interface is beneficial in three ways. First, it is easy to build query distribution and result aggregation on top of it and to extend those applications to handle new datatypes and aggregation functions. A node at the root of a tree issues a query to all other nodes by calling QTreeDown. When a node receives a QTreeUp, it optionally aggregates the attached message with the messages attached to other QTreeUp's it has received for that tree, and then delivers the aggregate to the parent.

Second, QTree relieves applications from the burden of reforming the tree topology in the face of node flux; QTree takes care of that, ensuring that a TreeId continues to refer to the tree rooted at a given node even in the face of node flux. We note that we have currently implemented reliability only in the TTREE configuration of QTree

Third, QTree enables experimentation with different spanning tree topologies; a new spanning tree implementation that maintains the QTree interface can immediately be substituted for an existing QTree implementation.

### 3.3. ISING

ISING, the Internet Sensor In-Network agGregator, is a simple query processor designed for continuous queries over streaming data received from PlanetLab-style sensors. We have built ISING on top of QTree as follows. A "root" ISING instance calls *NewTree()* to form a QTree. That ISING instance then activates its own sensor interface to receive queries from users. A user query is turned into a *QTreeDown* message that is sent down the tree, and aggregated results are sent back up the tree using *QTreeUp* messages. For this discussion we assume only one ISING tree exists in the system at any given time.

A user's query to ISING is a standard sensor query consisting of the following components.

- **sensor server port:** the port number of the sensor server, assumed to be running on that port on every node in the query tree.
- **sensor name:** the name of the sensor whose value the user wants returned from the specified sensor server.
- **host:** "ALL" if the query should be sent (using *QTreeDown)* to the indicated sensor server port on every node in the system, or the hostname of a single machine if the query should only be sent to one machine. In the latter case the query will be sent from the ISING root directly over TCP and the response returned directly over TCP.

- **aggregation operation:** one of the aggregation operations MIN, MAX, AVG, MEDIAN, SUM, and COUNT; or the special VALUE operator, which simply concatenates all values returned by the queried sensors.

- **epochDuration:** ISING supports two types of queries. (1) *Continuous* queries are issued once to the ISING root, and a new result tuple is delivered to the user at a fixed interval specified by the *epochDuration*. Note that the ISING root does not re-broadcast the query every *epochDuration* milliseconds; instead the query is registered locally with the ISING instance at each node. Then the local sensor is re-queried, and the result pushed up the aggregation tree, at that frequency. (2) *Snapshot* queries compute a one-time aggregate across the system. Such queries are particularly useful in problem diagnosis, when the user is exploring a decision tree of possible problem causes. A user specifies an *epochDuration* of zero to indicate that a query should be evaluated only once.

- **value selection criteria** *(optional)*: Some sensors may return multiple lines of output (rows), each with multiple fields (columns). The user may request that only rows in which a specified column number matches a specified regular expression are returned. The user may further request that only a specified column number from returned rows be returned.[1]

- **value predicate** *(optional):* For some queries a user may be interested only in sensor values provided by nodes that meet some other criteria. We therefore allow predicates to be applied to sensor requests. The semantics of these requests are as follows: the value(s) matching the `<sensor server host:port, sensor name, value selection criteria>` restrictions that would (in the absence of a value predicate) otherwise be considered "valid" are considered "invalid" if the specified value predicate is not met. A value predicate consists of one or more clauses, joined by AND or OR, that specify comparisons between any `<sensor server host:port, sensor name, value selection criteria>` tuple for any sensor on the machine where the original query is being processed, and a constant or another such tuple. The comparison operations supported are =, !=, >, <, >=, and <=.[2] "Invalid" values are ignored during aggregation.

While complicated queries are not easy to write by hand because they must be encoded in the URL that is sent to the ISING root, we expect that a program (such as ENTRIE), not a human, will be generating the URLs.

In response to a query with the above components, ISING returns one or more lines in comma-separated-value format, each line of the form `<sensor server host:port>`, `<timestamp when data item was generated`[3]`>`, `<data>`

As an aside, the ISING request semantics that we have described allow the user to specify aggregation over sensor data returned by a sensor running on the same port on every physical node in the system. This is not a problem for sensors that collect per-physical-node data, since there is no reason that there would be more than one sensor server for that sensor on a particular machine. But we also want to collect application-level data, and there may be more than one instance of an application running on a single physical node. For example, when evaluating the scalability of a decentralized routing layer, it is common to run more than one instance of the application on each physical machine to emulated a network with a larger number of nodes than there are physical nodes available for the experiment. In this case the ISING user will want to aggregate across *all* application processes running on *all* physical nodes. To address this need, we have implemented a sensor of which one instance runs on each physical node, that aggregates data from all instances of the sensor of interest on that node (one per application process of interest, each running a sensor server on a different port). This lower-level aggregator can be thought of as a recursive instance of ISING itself; it aggregates data from multiple sensors of interest with identical schemas on a physical node, and exports a single sensor interface to ISING, which queries it and aggregates across physical nodes.

Although we have described ISING as a query processor for sensors, it can be used identically to broadcast actuator invocations and aggregate their results (success/failure messages). This is possible because our actuators export a standard sensor interface. Actuators that we wish to activate simultaneously on every node in the system are ideally invoked through ISING. Examples that we have used include setting a simulated loss rate on each overlay link to some value to test the system's robustness to lost messages, and increasing the rate of workload generation uniformly across all nodes in the system. Of course, because actuators export a standard sensor interface, they can also be invoked directly rather than through ISING.

---

[1] For example, consider a query to a sensor that implements the Unix *finger* function. An intrusion detection application may be interested in querying this sensor at all nodes in the system to find out from where a particular user is logged on to those machines that she is using. This requires matching the username row(s) to the username and then extracting the "from" field(s) that indicates from where she is connecting

[2] For example, consider a query of the form "tell me which nodes are sending a lot of network traffic but have low CPU load"--these nodes might be launching a network attack. This query would be phrased such that the "sensor" would be a "tell me your hostname" sensor, while the value predicate would AND two predicates: the "amount of traffic sent over the last few minutes" sensor greater than some value, and the "CPU load" sensor less than some value.

[3] This timestamp is taken from the local clock on the machine running the sensor whose data is returned, so it is only comparable to other time values in the system if the clocks in the system are synchronized.

### 3.3.1. ISING failure handling

ISING should handle three types of failures: failstop failure of a sensor server or sensor, performance failure of an ISING instance (which may be due to performance failure of a sensor server or sensor), and failstop failure of an ISING instance (which may be due to failstop failure of a node).

The easiest of these failures to handle is failstop failure of a sensor server or sensor. This may happen, for example, because the sensor simply is not intended to run on some subset of machines, or because it has crashed and is now refusing HTTP connections. If an ISING instance detects this type of failure, it considers the value "invalid" for the purposes of aggregation.

An ISING instance may experience a performance failure if something has slowed down one of its children, the network between one of its children and itself, or a sensor server or sensor. To allow partial aggregates to be returned in the face of such failures, each ISING instance sets a timeout on receiving values from its children for each epoch. If the values for an epoch are not received from all children within the timeout interval, the aggregate of whatever data has been received thus far is passed up the tree as the value from that ISING instance for that epoch. When the values from the slow children for that epoch are received, they are discarded. In order to give timeouts a chance to propagate up the tree, each node's timeout should be proportional to the number of hops to that node's farthest descendant. As a heuristic, we set each node's timeout to

```
timeout = d * timeout_single
```
where
```
timeout_single = compute_max + latency_max
```
and *d* is the difference between the maximum anticipated depth of the tree and the depth of the node whose timeout value we are computing, $compute_{max}$ is the maximum amount of time we expect any node to need to compute the aggregate of its children's results after the last value is received, and $latency_{max}$ is the maximum anticipated one-way network latency between any two adjacent overlay nodes. This is a less precise version of the policy used in [8] for loss recovery.

Of course, whether a delayed value is useful to an ISING user depends on the *epochDuration* of the query -- a once-an-hour query whose response is delayed by two minutes would presumably be acceptable, while a once-a-minute query whose response is delayed by two minutes would be useless since another value will have been delivered in the interim.

QTree can hide the third type of failure, failstop failure of an ISING instance, from all other ISING instances: as nodes join and depart the system (voluntarily or due to failures), QTree re-forms the tree to incorporate the surviving nodes, assuming the root has not died. Our current ISING implementation uses QTree in a "static" mode, however, so that child death is treated identically to a performance failure. Thus until the child restarts, all messages from it and all of its descendants are lost, though results from the portion of the tree excluding the dead node and its descendants continue to be aggregated and returned to the ISING root. In addition to not re-forming the tree when a node dies, "static mode" prevents new nodes from joining the system. We are currently in the process of modifying ISING to exploit QTree's ability to dynamically re-form trees as nodes join and depart the system; until then, the same effect can be accomplished only by killing and restarting all ISING and QTree instances.

### 3.4. ENTRIE

Although ISING is a useful standalone service in its own right, its power is magnified when it is used as a data source for ENTRIE. ENTRIE is a configurable trigger system designed to issue queries to sensors (through ISING or directly), to continuously evaluate the results, and based on that evaluation to possibly invoke one or more actions. We built ENTRIE with two primary uses in mind: for running controlled experiments such as benchmarks and tests, and for performing system management.

Unlike ISING, which is designed as a single long-running service that can be accessed by anyone as part of the core system infrastructure, ENTRIE is designed to be instantiated separately by each person who uses it.

At ENTRIE's core is an XML configuration file specifying *triggers*, which are actions and the condition(s) that cause them to execute. This configuration file could be automatically generated from a more user-friendly syntax, but currently it is written by hand. ENTRIE's configuration language is intended to shield the user from such details as interacting directly with sensors and actuators, or having to write procedural specifications to evaluate triggers.

The core ENTRIE abstractions are *conditions* and *actions*. One or more conditions are associated with each action; when all of the conditions are met, the action is executed. The same conditions can be bound to different actions, to allow an "or" over conditions.

ENTRIE currently supports three types of conditions: *timer conditions*, *completion conditions*, and *sensor conditions*. *Timer conditions* specify that an action can be exe-cuted only after a certain time, and/or that an action cannot be executed after a certain time. *Completion conditions* specify that an action cannot be executed until another action(s) specified in the configuration file have completed. *Sensor conditions* specify the `<hostname:port number>` of a sensor server (which may be ISING), the name of the sensor of interest, the period with which the sensor should be queried (when querying through ISING, this is turned into the *epochDuration*), and one or more *data conditions*. A *data condition* specifies a condition over the value, or history of values, returned by the sensor. An example data condition is "the average load returned by the sensor over the last ten query periods is greater than 10." (If the sensor is ISING, then the loads that ENTRIE is averaging here are themselves the ten previous instantaneous average loads computed by ISING across all nodes in the system).

ENTRIE's actions invoke actuators. They can be named using the same syntax used for naming conditions, or they can use special syntax we have designed for starting, killing, or rapidly starting and killing ("churning") application processes to enable benchmarking. Actions may specify one of several *repeat* settings, to indicate that an action should be executed on the conditions' first transition from false to true, on every transition from false to true, periodically during the first true interval, or periodically during every true interval. The first type might be used to trigger a timer-based action in a benchmark, while the fourth might be used to page a system administrator periodically until a problem has been fixed.

Although ENTRIE represents a single point of failure for ACME, multiple ISING roots can be named as part of a sensor condition. The first ISING instance will be used as the default, and if the connection to it fails or times out, the next instance will be tried, and so on. Of course, if ACME or the node on which it is running fails, ACME will itself fail. When ACME is restarted it will re-read its configuration file and proceed as before, but having lost all of its history data. Standard "hot standby" replication of history data would allow ENTRIE to tolerate failures, but this functionality has not been implemented.

To illustrate ENTRIE's capabilities, we provide two concrete configuration examples here. Due to space constraints we are limited to fairly simple examples.

### 3.4.1. Example: benchmarking

Our first example would be used in benchmarking. First we start 150 nodes (instances of the application process). Fifteen minutes later we start a process of "churn" in which we start and kill new nodes repeatedly, such that the period between startups is drawn at random from an exponential distribution with mean 10 seconds, and the lifetime of each node is drawn at random from an exponential distribution with mean 30 seconds. Thirty minutes after churn has started, we end.

```
<action ID="1" name="startNode" timerName="T">
    <params numToStart="150"/>
    <conditions>
        <condition type="timer" value="0"/>
    </conditions>
</action>

<action ID="2" name="startNode" timerName="T">
    <params numToStart="1" distribution="exponential"
     randLifetime="true" meanLifetime="30000"/>
    <repeat distribution="exponential" randPeriod="true"
     meanPeriod="10000"/>
    <conditions>
        <condition type="timer" value="900000"/>
        <condition type="endDelay" value="1800000"/>
    </conditions>
</action>
```

### 3.4.2. Example: self-repair

Our second example, which might be used in system management, combines problem detection with a limited form of self-repair. We specify the following policy: every minute, if the load on the most highly loaded physical node is more than five times the maximum of the minute-by-minute average loads across the system during the past ten minutes, reboot that node. The user writing the file would replace text enclosed in [brackets] with a constant.

```
<action ID="1" name="EXECUTE" timerName="T">
    <params commandType="actuator"
     name="reboot"
     hosts="[ISING_host]:[ISING_port]"
     node="VARIABLE_host:[reboot_actuator_servr_port]"/>
    <conditions>
        <condition type="sensor" ID="systemAVG"
         name="load"
         hosts="[ISING_host]:[ISING_port]"
         node="ALL:[load_sensor_server_port]"
         period="60000" sensorAgg="AVG"
         histSize="10" histAgg="MAX" isSecondary="true"/>
        <condition type="sensor"
         name="load"
         hosts="[ISING_host]:[ISING_port]"
         node="ALL:[load_sensor_server_port]"
         period="60000" sensorAgg="MAX"
         histSize="1" operator=">"
         secondaryID="systemAVG" scalingFactor="5"/>
    </conditions>
</action>
```

## 4. System evaluation

In this section we evaluate ACME's performance, scalability, and robustness to overlay network message loss.

### 4.1. Experimental environment

We have tested ACME in three environments: a cluster, Emulab, and PlanetLab; we used the PlanetLab-compliant sensors we wrote as well as (when testing on PlanetLab) the PlanetLab *slicestat* sensor. We focused our evaluation efforts on Emulab [31], a large publicly-available cluster designed for distributed systems research, because it allows emulation of a realistic Internet topology without the resource contention and performance variability issues of PlanetLab. We have intentionally focused on evaluating our actual implementation, rather than simulating the system, in order to be sure to account for implementation details that could be overlooked in a simulation. While this decision limited our ability to evaluate our system's behavior for network sizes greater than 512, it does enhance our ability to claim that the system can be used, today, on wide-area distributed systems with hundreds of nodes, such as PlanetLab, Emulab, and large clusters.

For our Emulab experiments we created a topology according to the *transit-stub* model of GT-ITM [32]. We used three transit domains of approximately six nodes each, and about four nodes per stub domain. We used the latencies chosen by GT-ITM and verified that the resulting distribution of stub-to-stub ping times was similar to that observed on PlanetLab. We augmented the model with the following bandwidths: all stub-to-stub links are 100 Mb/s, all stub-to-transit links are 1.5 Mb/s, and all trans-to-transit links are 45 Mb/s. These values were chosen to model Fast Ethernet, T1, and T3 connections, respectively. The Emulab cluster's PCs are primarily 850 MHz Intel Pentium III's with 512MB of DRAM each.

QTree and ISING are written in Java and run (in the same JVM) under the IBM 1.3 JDK. ENTRIE is written in Java and runs in a separate JVM, under the Sun 1.4 JDK. In this section we have chosen two structured peer-to-peer overlay networks, Tapestry and Chord, as the applications that we use ACME to monitor and control. Tapestry is written Java, and we used the U.C. Berkeley Tapestry distribution downloaded on March 23, 2003 running on IBM JDK 1.3. MIT Chord is written in C++, and we used the distribution checked out of the Chord CVS repository on March 2, 2003. All code was run on the stub nodes in our topology, which were running Linux.

In this section we use the term *nodes* to refer to virtual nodes as opposed to physical nodes. We ran more virtual ISING/QTree nodes than we had physical nodes by running multiple instances of the ISING/QTree classes in the same JVM (across many physical machines). We ran up to three instances per physical node of the applications that we use ACME to monitor and control. In no experiment did we run more than one ISING/QTree process and three monitored-application processes per physical node, so we do not believe any of our measurements are influenced by abnormally high CPU load on physical nodes.

We use the DTREE and TTREE topologies of QTree. The TTREE topology is induced from the normal Tapestry network but with base 4 rather than base 16, to increase the number of levels in the induced aggregation tree. This causes the number of levels in the tree to scale with $\log_4(n)$ rather than the $\log_{16}(n)$, thus allowing us to evaluate TTREEs with average node depths of 6 using just 512 nodes (trees with average node depth 6 would require more than a hundred thousand nodes using base 16).

We evaluated one DTREE and one TTREE for each of the following network sizes: 64, 128, 256, 384, and 512 nodes. One stub node was selected at random from our network topology and used as the root of the trees in all experiments.

A node's depth in the tree is, by definition, the length of its routing path to the root. The maximum length is expected to be about $\log_4(n)$. However, surrogate routing [33] may add a small number of extra hops on that path towards the root, which increases the depth of a node. Thus we expect a network of size 512 to induce a tree with depth greater than $\log_4(512) = 4.5$. We find the average depth of nodes in our 512-node tree to be 6.6, confirming our expectation. This expectation was also confirmed for the other tree sizes used in our experiments.

### 4.2. Evaluating ISING

In this section we evaluate ISING's performance, scalability, and robustness.

Figure 2 plots the end-to-end time to issue a query to ISING and receive a response, as a function of aggregation operation and network size, for both the Tapestry-induced topology (TTREE) and the topology in which all nodes are connected directly to the root node (DTREE). We show only two aggregates, MIN and MEDIAN, because their performance behaviors are representative of *monotonic* and *non-monotonic* aggregates [15], respectively. MIN, MAX, COUNT, and SUM cause each node to pass exactly one value to its parent; AVG causes each node to pass exactly two values to its parent (the aggregated sum and the aggregated count); and MEDIAN and VALUE cause each node to pass to its parent all values collected from all of its children. Put another way, MIN, MAX, COUNT, SUM, and AVG benefit from aggregation, while MEDIAN and VALUE do not. To obtain these results we ran each query eleven times for each of the two topologies for networks of size 64, 128, 256, 384, and 512, and took the median response time for each query. This response time was measured as the delay between ISING receiving an HTTP request for a value and the result being delivered over that connection. MEDIAN was computed at each node by waiting for all children to respond, sorting the accumulated list of values, and selecting the middle value; while MIN was computed by recomputing a new MIN-seen-so-far each time a child value was received. In this experiment we used values generated internally by ISING at each node as the sensor value rather than having ISING at each node contact an external sensor, in order to isolate ISING's performance from that of specific sensors. The same node was the root of the tree for each network size.

The most important conclusion to be drawn from Figure 2 is that aggregation does help measurably. For a 512-node network, a true aggregate (TTREE-MIN) reduces latency by 76% compared to an aggregate that cannot reduce data as it flows through the overlay network (TTREE-MEDIAN), and it reduces latency by 54% compared to having all nodes send their values directly to the root using TCP (DTREE-MIN).

In analyzing these results we first discuss the relative latencies of the four configurations ({MIN, MEDIAN} x {TTREE, DTREE}), and then the relative slopes of each.

The absolute latencies we found are

*T-MEDIAN > D-MEDIAN = D-MIN > T-MIN*.

Our explanation of these latencies is as follows.

- TTREE-MEDIAN has higher latency than either DTREE operation because it ships the same amount of traffic over the bottleneck link (all values collected are sent over the root's network connection) and also ships additional data over other links (the links among non-root nodes) and incurs a delay proportional to the number of overlay hops between the farthest leaf and the root as parents at each level wait for their slowest child to complete.

- The two DTREE operations have approximately the same latencies because they ship identical amounts of data over exactly the same links. DTREE-MEDIAN is very slightly slower than DTREE-MIN because median cannot be computed until all child values are received, while MIN can be computed incrementally as child values are received.

- TTREE-MIN has lower latency than the DTREE operations because it ships the same total amount of traffic (each node sends one value) but the traffic is spread across many network links. A lesser effect that contributes to TTREE-MIN's performance is that the computation of the aggregate is overlapped among all nodes in the same level of the tree. The one drawback of TTREE, namely that it incurs a delay proportional to the number of hops between the farthest leaf and the root, apparently does not hurt performance as much as the load balancing of network traffic and computation helps it.

With respect to slope, the time to compute TTREE-MIN depends mainly on the depth of the tree, which we indeed found to be approximately constant across our tree sizes. When a new node is added, each existing node sends up the tree the same amount of data as it used to. The only extra work is that the new node's parent does one more unit of computation, and one new network message is shipped over the network link into the parent of the new node. The time to compute a TTREE-MEDIAN increases with a slope related to the depth of a the tree, because each new node increases by one unit the amount of network traffic sent along every overlay link on the path from the new node to the root. Finally, the slope of the DTREE lines are controlled by the fact that adding a new node increases by one message the amount of traffic sent over the heavily congested network link into the root.

Figure 3 plots the total number of bytes sent in response to a query as a function of aggregation operation and network size, for both the TTREE and DTREE topologies. Quite predictably, the DTREEs and TTREE-MIN all send exactly the same amount of data--every node sends one value, and the slope of the line is the number of bytes in a message. (Our messages are larger than they would be in a production system, as we include some debugging information; obviously the benefit from aggregation would be greater if messages were larger, and smaller if messages were smaller.) The TTREE-MEDIAN line can be understood as follows. Every node sends a number of message units equal to one more than the total number of its descendants. Therefore a new node causes *m* extra message units to be transferred, where *m* is the number of nodes on the path from the new node to the root. The average depth of a node expresses the average number of such intermediate nodes. Thus we expect the slope of the line to be approximately equal to the average node depth times the slope of a DTREE line. Indeed the slope of a DTREE line is about 100 bytes/node and the slope of the TTREE-MEDIAN line is about 600 bytes for node, for a ratio of about 6, which is about the average node depth we found for our trees.

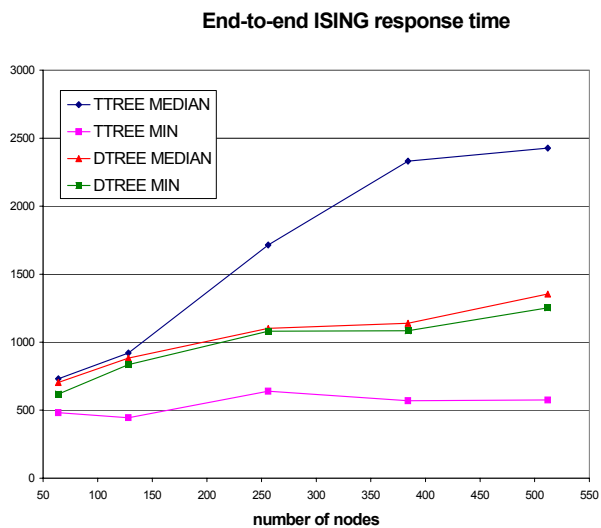

**Figure 2: ISING response time as a function of aggregation network size, topology, and operation.**

Finally, in order to investigate ISING's robustness to message loss, we instrumented ISING so that a fraction of QTreeUp messages would be dropped. In particular, each time a node is about to send a QTreeUp message to its parent, there is an *p*% chance that it will drop the message instead of sending it. Nodes decide to drop messages independently, based on a random number generator that is seeded differently on each node.

In an aggregation network there are two loss metrics of primary interest: the number of queries whose response incurs at least one loss, and the number of nodes partitioned from the tree by that loss(es). To assess these metrics, we recorded the values returned from a series of 100 COUNT queries issues by ISING; COUNT simply returns the number of nodes responding to a query. Table 1 shows, for a 512-node network, the total percentage of non-512 counts returned, representing the number of queries that experienced at least one message loss, and the average difference from 512 for non-512 counts, representing the number of nodes partitioned from the tree when there was loss. A full analysis of these results is not possible due to space constraints, but we make the following argument for their reasonableness. Assuming failures are independent, the expected fraction of queries that will return non-512 counts for loss probability $p$ is $1-((1-p)^{512})$, since in order for a 512-count to be returned, every link must not fail. (In a real network failures are not independent, but we leave exploration of a more realistic fault model to future work.) This expectation closely matches our findings in the second column. For the third column, in general, the higher in the tree a node is, the more of the tree that is lost when it, or its link to its parent, dies (at the two extremes, the root takes out everything, while a leaf only disconnects itself). But there aren't many nodes at the higher levels of the tree (only one root, but more than half the nodes are leaves). These two effects cancel each other out, and the expected number of nodes lost to a given failure is roughly proportional to the depth, which is roughly constant across all the cases. Also, as the loss probabilities increase, the probability of multiple losses in responding to a single query increases, which explains the increase in the average number of nodes lost as loss probability increases.

| Loss probability | % of lossy responses | average # of nodes lost |
|---|---|---|
| 0.01% | 4% | 5.25 |
| 0.05% | 17% | 5.71 |
| 0.10% | 42% | 6.22 |
| 0.15% | 48% | 7.82 |

**Table 1: Percent of query responses that lose at least one node's response, and average number of nodes lost for lossy responses, as a function of loss probability, for network size 512.**

### 4.3. Evaluating ENTRIE and ACME

Although ISING is an important part of ACME, we are also interested in the end-to-end performance of ACME: the time from a condition-satisfying value being produced by a sensor, until the action corresponding to the condition is invoked on the appropriate node(s). Assuming the action is to invoke an actuator on all nodes, this time is the sum of

(1) the time for a value to be received from a sensor
(2) the time for the aggregate value to reach the root
(3) the time for the root to pass the value to ENTRIE
(4) the time for ENTRIE to evaluate the trigger
(5) the time for ENTRIE to pass the actuator query to ISING
(6) the time for ISING to pass the actuator invocation down the tree
(7) the time to invoke the actuator.

The sum (6) + (2) is precisely the end-to-end number we measured in Section 4.2 (though in that case the "down" happened before the "up"). Due to time constraints, we were unable to evaluate ENTRIE's performance scalability, *i.e.,* the relationship between trigger time and such factors as total number of triggers, number of conditions associated with each trigger, and number of actions potentially triggered by the same condition. Indeed, the current version of ENTRIE was not designed with either performance or scalability in mind, but rather as a way to prototype our ideas about controlling distributed experiments and performing distributed system management using actuators. We did find that for a few triggers (actions), each with a few conditions, ENTRIE's trigger time never exceeded 100ms. In other words, (4)

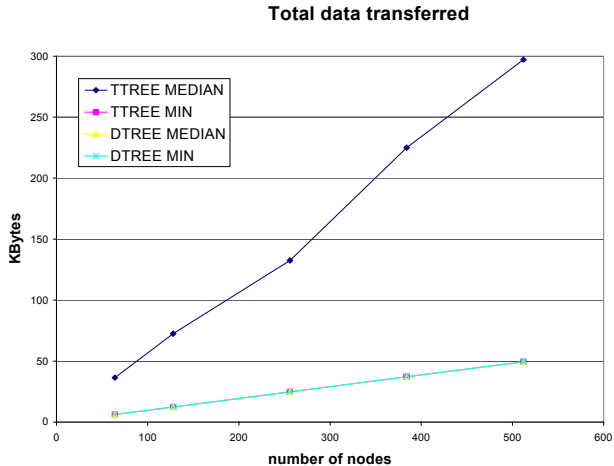

**Figure 3: Total bytes sent in computing an aggregate, as a function of aggregation network size, topology, and operation.** The three lower curves coincide.

was never more than 100ms.

The remaining factors in ACME's end-to-end performance are the latency to read from sensors (recall that the ISING root is itself a sensor, and that actuators are implemented as sensors). We did not consider the sensors and actuators used in our system to be performance critical (indeed, in a real system they are largely outside our control). We *were* interested, however, in evaluating the impact of our rather unorthodox decision to embed a small HTTP server inside every application instance for monitoring and control. In particular, we wanted to determine whether (i) reading from such a sensor contributes unduly to end-to-end overhead, and/or (ii) the extra work an application needs to do to handle HTTP requests significantly degrades the application's performance.

To answer (i), we simply measured the time it takes the HTTP server we embedded inside Tapestry to respond to a query about the number of messages it has routed since it was started. This took at most one second[1], which was also an upper bound on the latency for reading from other sensors we implemented in Java. Empirically the end-to-end performance for a condition becoming true at a sensor, to invoking an actuator on all nodes (the sum of 1-7 above), was found to always be less than four seconds when there is one instance of the queried sensor and actuator per physical node.

To answer (ii), we used ACME to benchmark a 100-node Tapestry network while querying every five seconds each Tapestry node's sensor for the number of Tapestry messages it had routed thus far. We set the ACME workload generator actuator to have each Tapestry node perform one *find_owner* lookup (as defined in [21]) every ten seconds. We measured for every one-minute interval: completion rate (percentage of lookups that do not time out), success rate (percentage of lookups that return the same mapping as the majority of the other lookups for the same identifier that were issued at about the same time), and mean latency for the lookups that completed. Figure 4 plots these metrics over time. Until minute 15 we do not query the Tapestry instances' internal sensors, and starting at minute 15 we query each sensor every five seconds. We see that completion rate and success rate are unaffected by the extra work and network traffic. The mean latency for minutes 6 through 15 (from once the system has stabilized until we start issuing sensor queries) is 329ms and the mean latency for minutes 16 through 30 is 351ms, an increase of 6.7%. We intentionally chose a relatively small network (100 nodes) to exaggerate the effect of querying the sensors; a larger configuration impacts each application instance the same, and because we use an aggregation sensor on each physical node to aggregate data from that node's Tapestry instances, the amount of network traffic would be the same if we ran more instances. Nonetheless, this one data point suggests that when using ACME to run benchmarks, it is wiser to log data to disk and use our log-reading sensor after the benchmark has run, rather than read benchmark statistics out of the application directly (at least using the application sensor as we have implemented it). Because of this overhead, the remaining graphs in this paper were collected by logging statistics to local disk and then aggregating the logs after the test was over.

---

[1] This latency is primarily due to the default settings that SEDA uses for polling its asynchronous sockets, which we did not attempt to adjust.

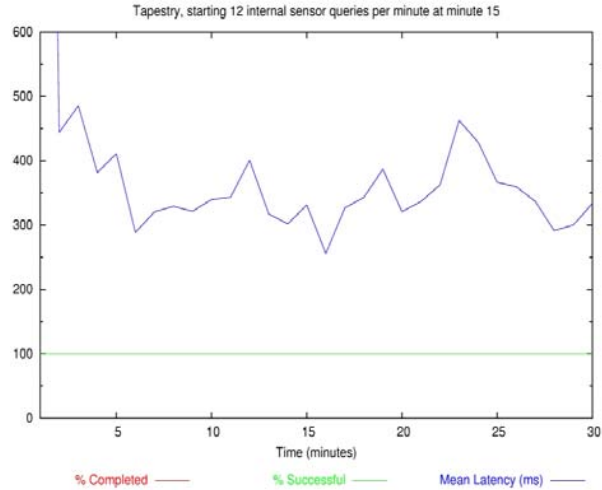

**Figure 4: Impact of querying Tapestry's application-embedded sensors every five seconds, 100 nodes.**

As we have mentioned, we believe that it is interesting to embed not only sensors, but also actuators, inside applications. Figure 5 and Figure 6 present graphs similar to Figure 4, this time demonstrating the impact of ACME invoking an application-embedded actuator at minute 15 and 16, respectively, to change the workload request rate from every twenty seconds to every five seconds. We have used a 150-node Tapestry/Chord network in this example to magnify the impact of increasing the workload. Figure 5 shows that Tapestry starts out with a less-than-100% completion rate due to timeouts, and that completion rate decreases after increasing the workload. Also, mean response latency increases significantly after minute 15. Figure 6 shows that shows that Chord maintains a 100% completion rate and its mean latency is less affected than is Tapestry's.

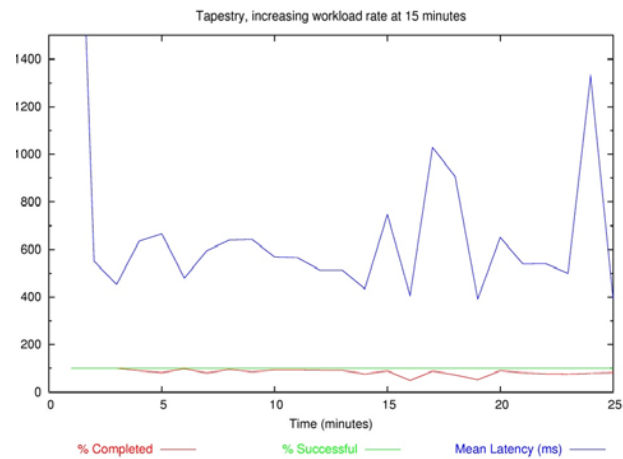

**Figure 5: Impact of quadrupling Tapestry request rate using application-embedded actuator, 150 nodes.**

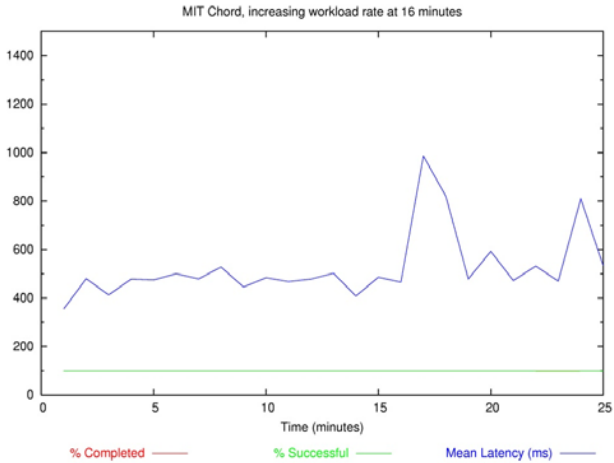

**Figure 6: Impact of quadrupling Chord request rate using application-embedded actuator, 150 nodes.**

We have used ACME to generate a large number of other scenarios, including scenarios described in [21] and [33], but due to space constraints we do not present their graphs here. Finally, note that our results from measuring Tapestry and Chord are not comparable to those presented in those papers because we used different versions of the software, different numbers of nodes, and Emulab instead of PlanetLab.

## 5. Related work

ACME's goals of providing an infrastructure for monitoring, analyzing, and controlling Internet-scale systems are precisely those of the recently-proposed Internet "knowledge plane" [7]. In terms of existing systems, ACME as a whole is most similar to Sophia [29], which in turn builds upon InfoSpect [22]. In a sense, ACME takes the opposite philosophy of Sophia; ACME provides a very constrained query and trigger language, and a much smaller implementation, for accomplishing similar tasks. An analysis of the tradeoffs between the two systems in terms of expressiveness, performance, robustness, and resource consumption, is left for future work.

A number of systems have recently been developed to query Internet-distributed data, making them closely related to ISING. PIER [11] is a relational query engine originally design to efficiently query data stored on nodes in a DHT, and that has recently been expanded to query PlanetLab sensors directly. In comparison, ISING is a less general query engine, but it supports continuous queries and hierarchical aggregation. Astrolabe [28] is also a relational query engine for Internet-scale systems, designed largely for distributed system monitoring. Like ISING it performs hierarchical aggregation, but Astrolabe's hierarchy is based on pre-specified administrative domains rather than a structured peer-to-peer overlay network's self-organizing topology, and data is disseminated using peer-to-peer gossip rather than ISING's more structured tree-based communication. IrisNet [18] is a distributed XML-based query engine for Internet-distributed multimedia sensors; it uses distributed filtering and hierarchical caching, distinguishes between sensing nodes and query processing nodes, and uses direct network connections rather than an overlay network. Netbait [6] is a distributed worm detection service that allows users to query worm signature data stored in local databases on Internet nodes; queries are distributed, and the results returned, using a mechanism almost identical to TTREE. Netbait uses one hard-coded aggregation operation (concatenate children's data) as results flow up the tree. Netbait is a good example of an application that could be built on top of ISING, assuming the existence of a sensor interface to the local node databases. Finally, Ganglia [9] is a distributed monitoring system that uses IP multicast to collect monitoring data within a cluster and polling over statically-configured TCP connections to collect data from each cluster to a centralized monitoring node. Compared to ISING, Ganglia's using direct connections instead of an overlay, and not performing wide-area aggregation, limit its scalability.

ISING's data aggregation over Internet sensors bears a strong resemblance to data aggregation in wireless sensors networks [14] [15], though the motivation for aggregation in those systems is primarily energy savings as opposed to performance and wide-area bandwidth reduction. Indeed, ISING is in many ways a reflection of TAG [15] onto the Internet, but with a more constrained query syntax. [3] describes aggregation as a possible application of Ephemeral State Processing.

QTree's query broadcast is related to application-level multicast, an area with a rich literature. A number of systems provide this service by building upon overlay networks, including [4] [5] [12] [13] [20] [34]; indeed, [34] is built on Tapestry, though it exists only as a simulation.

Finally, much recent work has focused on benchmarking and testing systems by measuring attributes such as performance under faults; this is one important application of ACME. [21] and [17] describe performance and performability benchmarks, respectively, for peer-to-peer routing layers and cluster software, respectively. Each built its own *ad hoc* monitoring and fault injection infrastructure, something for which ACME provides reusable building blocks. Finally, ACME bears some resemblance to NFTAPE, a tool for constructing fault injection experiments for small-scale distributed systems [26]. Unlike

NFTAPE, ACME is designed to scale to Internet-scale systems and uses a sensor/actuator interface to communicate with monitoring and fault injection components.

## 6. Future work and deployment

We are interested in enhancing ACME's performance, robustness, and functionality in a number of ways, while maintaining the application-specific focus that sets ACME apart from general-purpose distributed query processors and distributed programming environments.

First, we intend to evaluate additional QTree overlay topologies. In contrast to wireless sensor networks, where nodes can only route directly to other nodes within radio distance, Internet aggregation networks can form any overlay topology, because any node can route to any other node over IP. Thus we see a wide opportunity to investigate the performance and robustness of a host of hierarchical aggregation networks, including ones that are derived from structured peer-to-peer overlay networks, ones that are based on unstructured networks, and ones that are derived from social structures (*e.g.,* based on administrative domains), particularly in the face of real-world failure modes and queries that might be scoped based on geographic distance, administrative domain hierarchies, or network distance. Less structured data dissemination protocols, such as gossip, are also of interest [10].

Within ISING, we would like to investigate the potential performance improvement from caching values at the ISING root and non-root instances, as well as the sharing of query subexpressions (*i.e.,* an individual or aggregate sensor value). Also, for some applications, sampling a fraction of the sensors on each epoch may improve performance without significantly sacrificing data quality. We also intend to add additional aggregation functions such as COUNT DISTINCT and HISTOGRAM, and to investigate allowing user-defined aggregation functions specified within ISING queries as URL pointers to custom aggregation code. Finally, we would like to implement a mechanism for explicitly notifying the issuer of a query when she is receiving a partial aggregate due to timeouts, as opposed to a complete aggregate for which all nodes responded in a timely fashion.

From a more practical standpoint, we intend to integrate QTree and ISING into a simulation framework that will allow us to evaluate performance beyond the 512 virtual nodes to which we were limited for this paper by virtue of evaluating only a real implementation. Also, we would like to use ACME to monitor and control additional applications beyond Tapestry and Chord. Finally, we intend to add support for "streaming sensors," *i.e.,* sensors that return a new tuple of data periodically over a persistent connection to an ISING instance. This raises interesting issues related to matching the user's *epochDuration* to the rate at which new data is supplied by the sensor.

Finally, we would like to expand ENTRIE's functionality in four directions. First, we would like to add a layer of syntactic sugar on top of the current XML configuration file, particularly in the hopes of developing a general language capable of expressing the full range of fault injection actions and other control actions that benchmarkers, testers, and service operators might need. Second, we would like to add new sensors and actuators to increase the range of conditions and actions that can be utilized. Much longer term, we would like to provide ENTRIE as a service; users should be able to dynamically add and remove triggers stored on, and executed by, an "ENTRIE server." Such a service brings up a host of protection and security issues which must be considered. A final long-term direction for ENTRIE is to exploit statistical anomaly detection techniques over monitoring data to automatically instantiate, or to suggest to an operator, conditions that should trigger actions such as recovery from failures, quarantine of security problems, or operator notification for manual intervention. For this and other operations that might require large amounts of historical monitoring data, storing metrics on disk in raw or aggregate form, at the ISING root and/or non-root ISING instances, may be necessary.

We intend to deploy ISING as a continuously-running service on PlanetLab soon.

## 7. Conclusion

In this paper we have described ACME, a flexible infrastructure for Internet-scale monitoring, analysis, and control in support of activities such as benchmarking, testing, and self-management. Users create triggers using XML; one possible source of data for these triggers' conditions is ISING, a simple distributed query processor that broadcasts queries to, and aggregates data streams derived from, PlanetLab-style sensors. ISING can also be used as a sink for the triggers' actions, which is particularly useful when a trigger must invoke an actuator on all nodes in the system. ISING is in turn built on top of QTree, which imposes a uniform query/response interface on top of various overlay network configurations.

In evaluating ISING's performance and scalability, we found that for one 512-node system running atop an emulated Internet topology, ISING's use of in-network aggregation over a spanning tree topology derived from the Tapestry structured peer-to-peer overlay network reduced end-to-end query-response latency by more than 50% compared to using direct network connections or the same overlay network without aggregation. We also found that an untuned implementation of ACME can invoke an actuator on one or all nodes in response to a discrete or aggregate event in less than four seconds. Finally, we demonstrate ACME's ability to monitor and benchmark peer-to-peer overlay applications. To accomplish this we have written sensors for measuring application-level behavior and actuators for generating perturbations such as starting and killing process and nodes, varying the applied workload, varying emulated network behavior, and injecting application-specific faults.

ACME is just a first step in investigating the issues related to building an infrastructure for comprehensively understanding, testing, and managing Internet-scale applications. We look forward to future work in this area by ourselves and others.